\documentclass[]{spie}  %>>> use for US letter paper
\usepackage{amsmath,amsfonts,amssymb}
\usepackage{graphicx}
\usepackage[colorlinks=true, allcolors=blue]{hyperref}
\usepackage[T1]{fontenc}

\usepackage{siunitx}
\newcommand{\SIadj}[2]{\SI[number-unit-product={\text{-}}]{#1}{#2}}

\usepackage{xcolor}

\title{Design, assembly, and initial test results of a cryostat for holographic characterization of microwave telescopes}

\author[a,b]{Jon E. Gudmundsson}
\author[a]{Miranda Eiben}
\author[a]{Thomas J.L.J. Gascard}
\author[c]{Eve M. Vavagiakis}
\affil[a]{Science Institute, University of Iceland, Sæmundargata 2, 102 Reykjavík, Iceland}
\affil[b]{The Oskar Klein Centre, Department of Physics, Stockholm University,
AlbaNova, SE-10691 Stockholm, Sweden}
\affil[c]{Department of Physics, Duke University, Durham, NC 27710, USA}
\authorinfo{Further author information: Send correspondence to JEG\\JEG: E-mail: jegudmunds@hi.is}

\begin{document} 
\maketitle

\begin{abstract}
We describe the design, fabrication, assembly, and room-temperature vacuum qualification of a \SIadj{1.4}{\meter} long cylindrical cryostat developed for holographic testing of cryogenic microwave telescope optics. The system consists of a welded 6061-aluminum vacuum vessel containing nested 45- and \SIadj{4}{\kelvin} aluminum radiation shields, cooled by a two-stage pulse-tube cryocooler through commercial OFHC copper flexible thermal straps. The intermediate \SIadj{45}{\kelvin} stage intercepts radiative, conductive, and wiring heat loads from room temperature, while the \SIadj{4}{\kelvin} stage defines the volume used for optical testing. The cryostat includes a \SIadj{38}{\centi\meter} aperture for a microwave-transparent vacuum window and is sized to accommodate full-scale optical assemblies relevant to cosmic microwave background instrumentation. We summarize the cryostat architecture, lightweighted radiation shields, G-10 support flexures, welded vacuum-vessel fabrication, and room-temperature leak-checking campaign. Iterative helium leak checking and weld repair reduced the observed leak rate in the cryostat by over three orders of magnitude.

\end{abstract}

% Include a list of keywords after the abstract 
\keywords{instruments, cryostat, vacuum vessel, cosmology: cosmic microwave background}

\section{Introduction}
\label{sec:intro}  
Modern millimeter-wave telescopes used to study the cosmic microwave background (CMB) rely heavily on cryogenic infrastructure. Thermal radiation from warm surfaces can contribute directly to the optical loading of detectors and introduce systematic effects that complicate the interpretation of the measured sky signal. For this reason, the design of CMB instrumentation is inseparable from the design of the cryogenic architecture. This proceeding describes the design, construction, testing, and assembly of a dedicated cryostat developed for the characterization of microwave telescopes. The system will support precision measurements of optical hardware at cryogenic temperature and provide a flexible platform for validating telescope performance under conditions representative of modern CMB experiments.

%%%%%%%%%%%%%%%%%%
\section{Design requirements and cryostat architecture}
\subsection{Overall geometry and optical volume}

\begin{figure}[t!]
    \centering
    \includegraphics[width=0.8\textwidth]{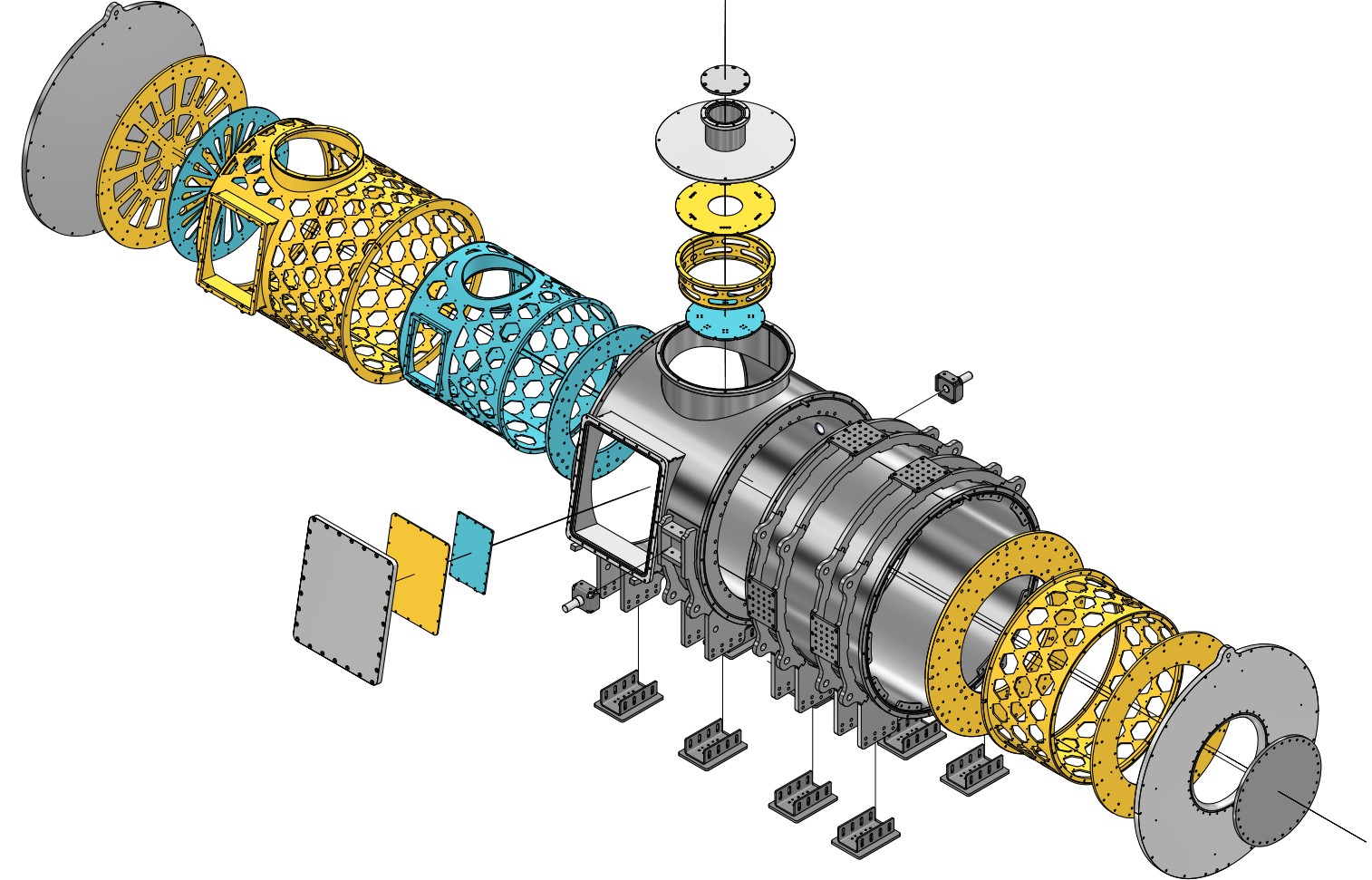}
    \caption{Exploded view of the cryostat showing the vacuum vessel in gray and the 45- and \SIadj{4}{\kelvin} shells colored in yellow and cyan, respectively.}
    \label{fig:exploded_view}
\end{figure}

The cryostat represents a critical part of the ERC-funded CMBeam research project. The project proposal, which was written in 2021, pointed out that the CMB community would benefit from further development of cryogenic holography measurement methodologies, following influential work described in publications such as Takakura et al.\ (2019), Davis et al.\ (2019), and Chesmore et al.\ (2022) \cite{Takakura2019, Davis2019, Chesmore2022}. To facilitate such efforts, a cryostat capable of housing a full-size optical system for a CMB experiment was required. The CMBeam project began in earnest in the summer of 2023.

The initial concept for the cryostat is based on the Mod-Cam cryostat \cite{Vavagiakis2022}. Early work on the full cryostat design was led by Dr.\ Thomas Gascard following the development of a fully parametrized COMSOL thermal model \cite{Gascard2024}. This model was used to study design that varied in size and in implementation of thermally insulating G-10 support flexures. A year after the completion of that work, members of our research group approached a number of US and European machining companies with experience in the fabrication of cryostats. We also approached local machining companies in Iceland and had fruitful initial discussions with engineers at Hedinn \cite{Hedinn}. In the end, we concluded that the prospect of collaborating on the first cryogenic system to be built in Iceland outweighed the potential risks associated with fabrication challenges stemming from lack of experience with these kinds of systems. Figure~\ref{fig:exploded_view} shows the final version of the mechanical design which was approved for machining in late October 2025, just four months after first verbal agreement on the construction of the cryostat was made. The design includes a \SIadj{38}{\centi\meter} aperture that allows for the installation of a transparent vacuum window.

The cryostat is composed of a cylindrical vacuum vessel shell that houses two nested radiation shields (see Figure~\ref{fig:exploded_view}). The vacuum vessel has a total volume that slightly exceeds $\pi \times (\SI{0.464}{\meter})^2  \times \SI{1.352}{\meter} = \SI{914}{\liter}$ due to the additional volume of the cryocooler chimney and RF plates that protrude slightly outwards. The intermediate \SIadj{45}{\kelvin} temperature stage reduces radiative loading from room temperature and intercepts conductive heat through G-10 supports as well as electrical wiring. The \SIadj{4}{\kelvin} shield defines the cold experimental environment, which has a volume of $\pi \times (\SI{0.302}{\meter})^2  \times \SI{0.576}{\meter} = \SI{165}{\liter}$; although it is expected that microwave optics tubes that are mounted inside the \SIadj{4}{\kelvin} stage will protrude outwards into the \SIadj{45}{\kelvin} region with additional radiative shielding. The mass of the vacuum vessel, \SIadj{45}{\kelvin} stage, and \SIadj{4}{\kelvin} stage is 316, 78, and \SI{25}{\kilo\gram}, respectively.

\subsection{Cryocooler and thermal interfaces}
\begin{figure}[t!]
    \centering
    \includegraphics[width=0.8\textwidth]{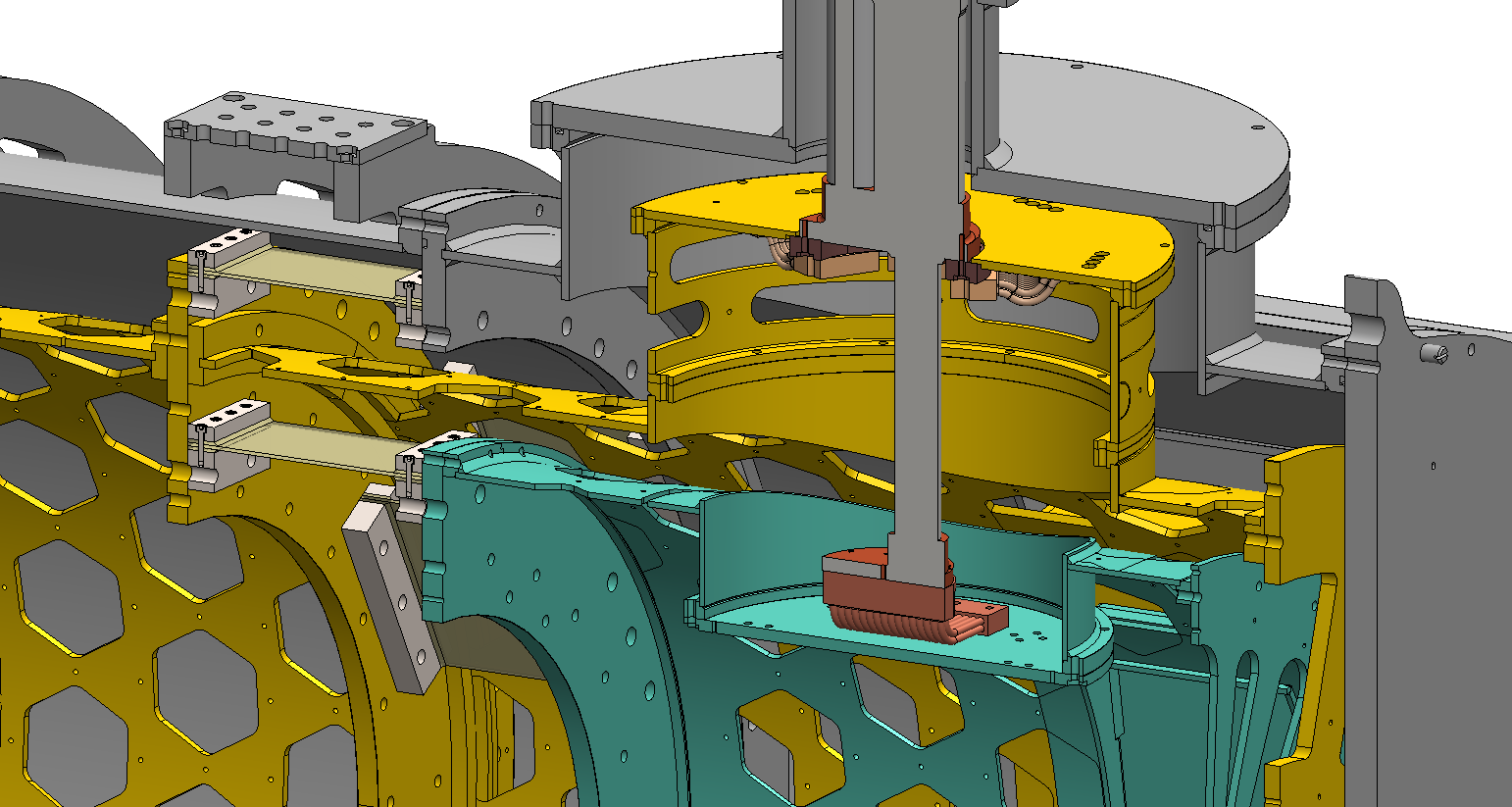}
    \caption{Heat straps are bolted from cryocooler thermal stages into circular interface plates  of the 45- and \SIadj{4}{\kelvin} stages. Six heat straps are symmetrically distributed around the first stage and connected to the top plate.}
    \label{fig:heat_straps}
\end{figure}

The system will be cooled using a Cryomech PT420 cryocooler from Bluefors (already procured) which provides a nominal cooling power of 55 and \SI{2}{\watt} at 45 and \SI{4}{\kelvin}, respectively.\cite{bluefors} Thermal links between the cryocooler cold head and the aluminum radiation shields are provided by commercial OFHC copper flexible thermal straps manufactured by Technology Applications, Inc (see Figure \ref{fig:heat_straps}) \cite{tai_thermal_straps}. The design uses six P50-502 CuTS straps to couple the first-stage cold head to the \SIadj{45}{\kelvin} shield and two CS-94A CuTS straps to couple the second-stage cold head to the \SI{4}{\kelvin} stage. The straps are fabricated from C10100 OFHC copper braid and end fittings, providing high thermal conductance while maintaining mechanical compliance between the pulse-tube cold head and the relatively large shield structures. 

The supplier estimates an end-to-end thermal conductance of \SI{2.84}{\watt\per\kelvin} per strap at 45 K, corresponding to a combined conductance of \SI{17.0}{\watt\per\kelvin} for the six-strap assembly. This conductance estimate corresponds to the installed strap geometry and includes the expected contribution from the mechanical interfaces. The supplier estimated an end-to-end thermal conductance of \SI{1.43}{\watt\per\kelvin} per strap at \SI{4}{\kelvin}, giving a combined conductance of \SI{2.86}{\watt\per\kelvin} for the two-strap assembly. 

\subsection{Radiation shields and MLI}

We employ a two-stage aluminum 1050 alloy radiation shield structure with repeating hexagonal lightweighting patterns to reduce mass. This lightweighting technique will also be used on the cryostat for the Taurus balloon-borne experiment \cite{May024, Tartakovsky2024, May2026}. The 45- and \SIadj{4}{\kelvin} radiation shields are \SIadj{4}{\milli\metre} thick and weigh approximately 78 and \SI{25}{\kilo\gram}, respectively. The \SIadj{87}{\milli\metre} diameter hexagonal cutouts are removed during watercutting prior to rolling of the shield structure. During the watercut process, small, roughly \SIadj{2}{\milli\metre} diameter, holes are watercut to facilitate potential mounting of thermometers or other electronics (see Figure~\ref{fig:heat_straps}).

Both radiation shields will be covered on the outside with 20 sheets of PolaWrap Al-PET-Al multilayer insulation (MLI) supplied by 
Aerospace Fabrication \& Materials, LLC.\cite{aerospace_fabrication_contact} Thick aluminum foil, intended to reduce radiative loading from the bottom layer of MLI blanket and improve conduction to the blanket, will be wrapped around the radiation shields and taped down before installing the MLI. The MLI, which provides the primary reduction of radiative exchange between stages, will be loosely attached to the aluminum shields at variation locations using nylon string. Blankets will be cut to appropriate sizes based on cardboard templates that have been fitted to the fabricated shields. The lightweight aluminum geometry is intended to reduce thermal mass and therefore speed up cooldown while also reducing the mechanical load on the G-10 flexures that support the inside of the cryostat. We have not performed detailed thermal simulations that probe the impact of this feature on cooldown times and overall thermal performance. The lightweighting also reduces in-plane thermal conductance, so temperature gradients will be evaluated during initial cooldown. 

\subsection{Thermally isolating flexures}
\begin{figure}[t!]
    \centering
    \includegraphics[width=0.3\textwidth]{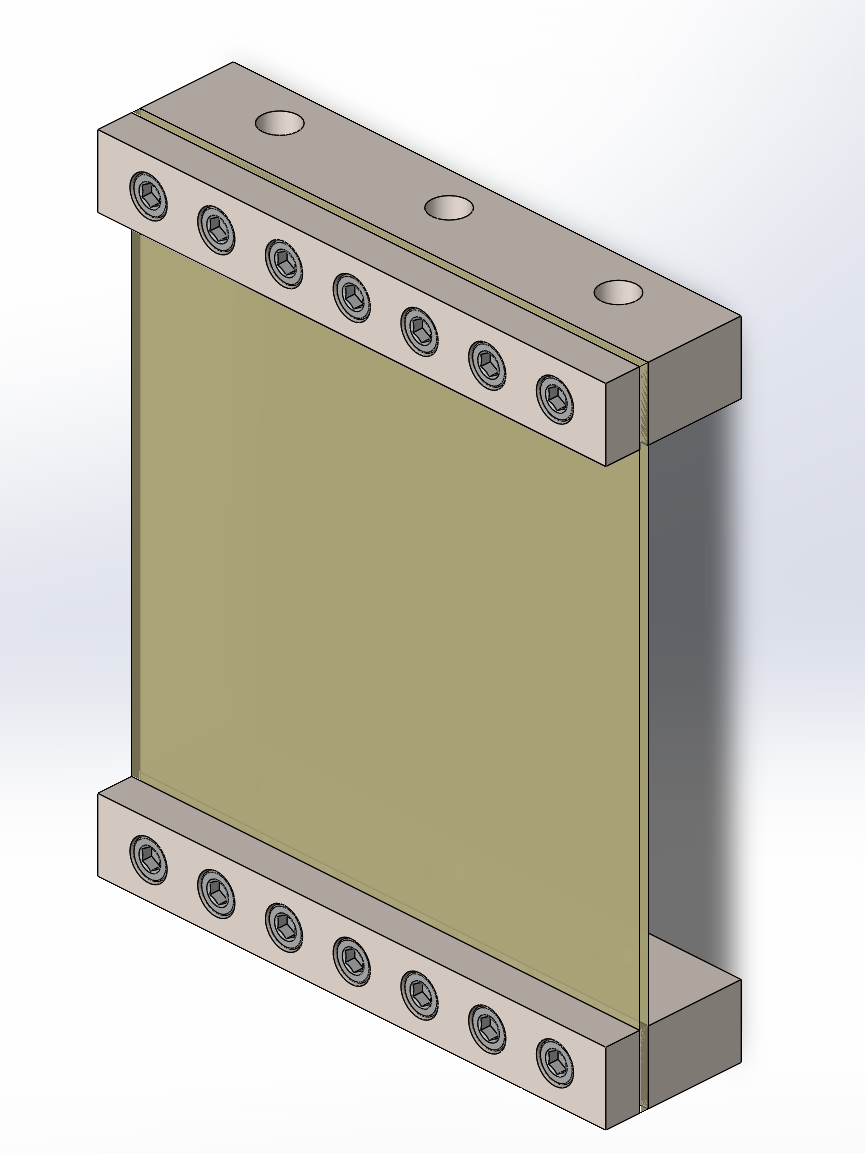}
    \caption{G-10 flexure design used for both temperature stages. A \SIadj{2.4}{\milli\meter} thick G-10 plate sandwiched between two stainless steel blocks using M6 socket head screws. }
    \label{fig:G10}
\end{figure}

The two temperature stages are each supported by six thermally isolating G-10 flexures (see Figure~\ref{fig:G10}) that are evenly distributed around the symmetry axis of the cryostat. These flexures are \SI{2.4}{\milli\meter} thick and have a $15 \times \SI{18}{\centi\meter}$ cross section. They are clamped (not glued) between stainless steel blocks using seven M6 socket head screws on each side \cite{Galitzki2024,Tartakovsky2024}. We rely on tight, \SI{6}{\milli\meter}, clearance holes in the G-10 to reduce risk of significant slippage if the bolt connections begin to fail. Should early testing of the cryostat performance suggest that these flexures are underperforming, we will modify the flexure to combine Loctite EA 9309 epoxy with mechanically constraining pins.\cite{silmid} 

Once clamped, the G-10 cross section that is not touching stainless steel is $15 \times \SI{14}{\centi\meter}$. We use the same flexure design for the two thermal stages, but the forces sustained by the flexure depend strongly on their orientation since the cryostat is designed to primarily rest in a horizontal configuration. Room temperature mechanical testing of the design, which included, tension, compression, and torsion, gives mechanical properties that are consistent with the literature\cite{Runyan2008} and suggests that a set of six flexures will readily support \SI{160}{\kilo\gram} of mass --- the total weight of radiation shields and optical components suspended of the vacuum vessel.

\begin{figure}[t!]
    \centering
    \includegraphics[width=0.5\textwidth]{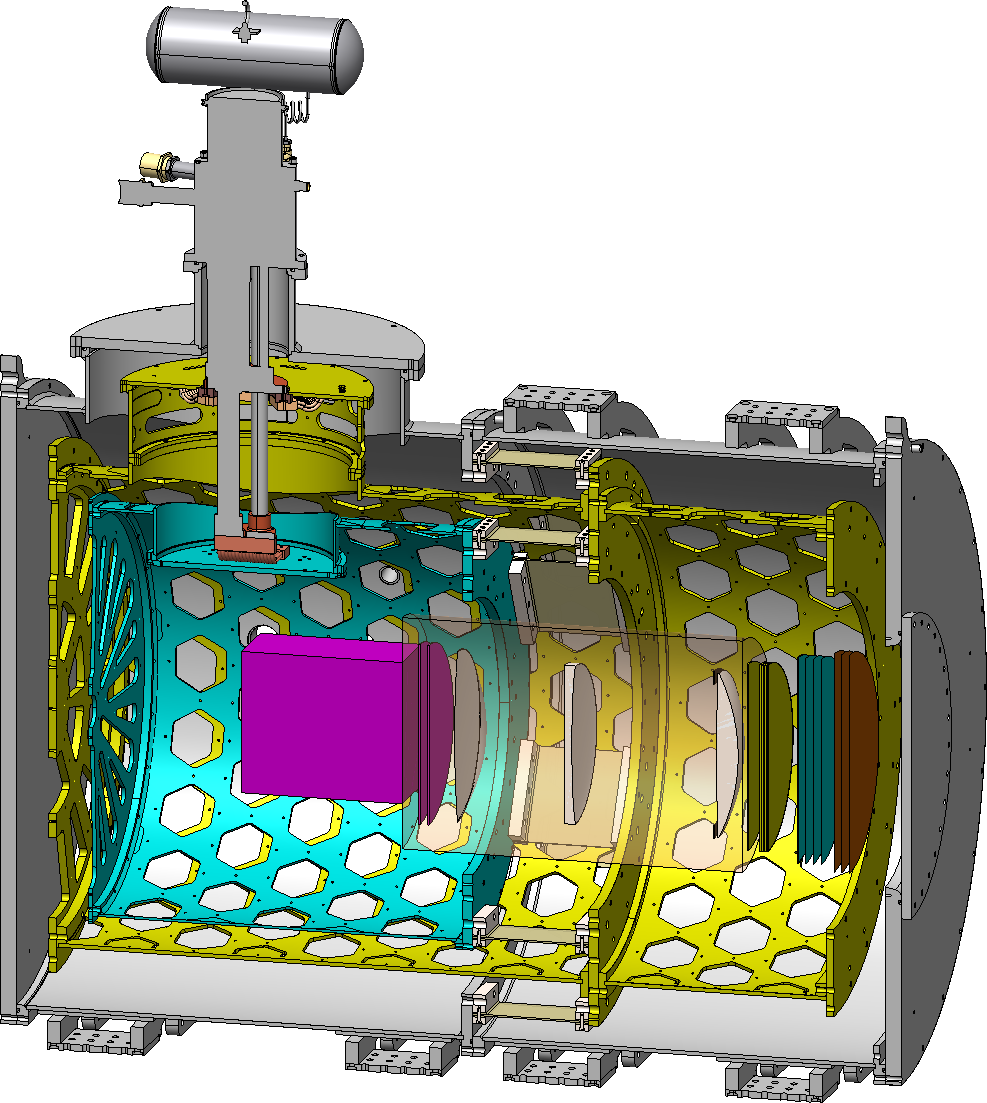}
    \caption{A three-lens refracting telescope design for the Taurus experiment, associated filters and focal plane hardware (pink) shown inside the cryostat. The total length of the optics tube from the back of the focal plane proxy to the front of the filters is approximately \SI{790}{\milli\meter}}.
    \label{fig:cryostat_woptics}
\end{figure}

The six flexures are bolted to circular flanges on the cryostat with three M12 bolts (see Figure~\ref{fig:heat_straps}). Both sets of flexures are located near the center of the cryostat (see Figure~\ref{fig:cryostat_woptics}). We estimate that each flexure will conduct \SI{270}{\milli\watt} from 300 to \SI{45}{\kelvin} and \SI{18}{\milli\watt} from 45 to \SI{4}{\kelvin}.

\subsection{Component-level fabrication and Welding}

The whole vacuum vessel (\SIadj{300}{\kelvin} shell), which weighs approximately \SI{316}{\kilo\gram}, was built from parts that were originally watercut from three \SIadj{20}{\milli\metre} and two \SIadj{25}{\milli\metre} thick aluminum 6061 sheets of size $1.0 \times \SI{1.5}{\meter}$. In order to save material cost, the four annular flanges that make up the top, center ($2\times$), and bottom structural supports of the vacuum vessel were assembled from three \SI{120}{\degree} circular segments that were welded together radially to form a full circle. Imperfections in these radial welds mean that leaks have a particularly high likelihood of forming (see Section~\ref{sec:leak_checking}).

Two mounting points on the side of the vacuum vessel, near the center of mass, allow for the installation of trunnions that facilitate elevation rotation should that need arise in the future. The vacuum vessel design includes eight structural support flanges that are spot welded to the vacuum vessel shell. Four of those flanges, those on the top cylinder, extend around the full circle of the vacuum vessel while four are truncated by the trunnion support mechanisms and an RF plate. The front structural flanges include anchor points for six multi-purpose mounting plates on the sides and the top of the cryostat. The flat RF plate facilitates the mounting of hermetic electrical connections (temperatures, heaters, and coaxial cables). Eight legs are bolted to the structural support flanges with slotted clearance holes that enable leveling of the cryostat relative to a slotted lifting table supplied by Siegmund with \SI{1000}{\kilo\gram} lift capacity.

The side of the bottom cylinder that is not visible in Figure~\ref{fig:exploded_view} includes five aluminum pumpout ports, supplied by Atlas technologies\cite{Atlas}, two KF50 ports and three KF25 ports. One of the KF50 ports will be used for pumping while the remaining ports serve will be used as backup, for pressure relief valves, or for electronic feedthroughs.

\section{Room temperature vacuum qualification}
\label{sec:leak_checking}
The leak checking of the vacuum vessel began in March 2026 and lasted until May 2026. During this time period, members of the research group would visit Hedinn a total of 11 times to pump down the cryostat and perform leak checking. The group benefited immensely from being able to work on the cryostat in a high-bay facility with access to a jib crane for easy assembly work. Two individuals performed the aluminum vacuum welds for this cryostat. We found that welds performed by the more experienced welder had a lower likelihood of failing. 

The first pumpdown in March showed clear signs of leaks in the vacuum vessel. Initial measurements gave an equilibrium leak rate of \SI{5e-6}{\milli\bar\liter\per\second} as read out by our Pfeiffer ASM340 leak checker. A relatively high equilibrium pressure of \SI{0.4}{\milli\bar} while pumping with a Triscroll 600 told the same story; the asymptotic pressure with the vacuum vessel valved off is \SI{3e-3}{\milli\bar}. We also implemented a leak-up test that consisted of valving off the cryostat and reading out the pressure on our pressure gauge before and after the 10-min integration time. The first leak-up test showed a pressure rise of \SI{0.1}{\milli\bar\per\minute}. Following this disheartening result, we embarked on a time-consuming leak checking campaign that lasted two months. 

During this period, we identified leaks in four key locations of the vacuum vessel: the circular flanges, the larger of the two circular cryocooler flanges (chimney), the welds on the KF pump out ports, and the welds around the rectangular RF plate. We were typically able to pinpoint those leaks down to a 1- or \SIadj{2}{\centi\meter} long section of a weld with careful helium spritzing in combination with two techniques described below.

We used non-hardening Butyl caulk (McMaster: 9408T15) that we could press over suspect welds. When the caulk was over a leak, we could both see a reduction in the equilibrium pressure and a drop in the response to helium spritzing. Using trial and error, we were usually able to localize leaks down to relatively small areas which were then cut out using a pneumatic drill equipped with a reamer before TIG rewelding with ER4043 weld wire. 

In some cases, we also found that spraying suspect areas with isopropanol or acetone would produce repeatable signals on the Pfeiffer MPT 200, digital Pirani cold cathode gauge installed on the pumping manifold. We could only see this when the system had reached an equilibrium pressure of order \SI{0.1}{\milli\bar}. For example, at a certain time during our leak checking effort, the pressure gauge showed a steady pressure of \SI{0.227}{\milli\bar} (not changing at the 3rd significant digit on few-minute time scales). We could then spritz the suspect area with these solvents and see a jump in pressure up to \SI{0.231}{\milli\bar}. The pressure would then fall back down to the original number in approximately one minute. The tests were repeatable which indicated that some liquid was reaching the inside of the vacuum vessel, evaporating, and subsequently impacting the pressure gauge.

The most challenging leaks that we identified involved flanges that had been fabricated from three segments. By construction, failures in those welds were radial which meant that gas had a relatively direct path into the vacuum vessel. In one case, we found a crack that was clearly visible by eye passing through an O-ring groove on the chimney for the cryocooler. In order to fix this crack, we decided to weld up the whole \SIadj{500}{\milli\meter} long O-ring groove and re-machine it. This was a lengthy process since the flange had to be re-planed after the repair. None of the other repairs required re-machining.

The final leak check prior to shipping of the cryostat to the University of Iceland took place in late May. A \SIadj{10}{\minute} leak-up test showed a pressure rise of \SI{3.61e-3}{\milli\bar\per\minute} which corresponds to roughly \SI{0.5}{\milli\bar} per day. A pressure of this magnitude can be explained by residual outgassing from oils in the vacuum vessel or O-ring permeability. Our leak checker at this point showed \SI{7.0e-9}{\milli\bar\liter\per\second} while the equilibrium pressure read out on the manifold was \SI{7.12}{\milli\bar}; only a factor of two higher than the pressure read out by the pressure gauge when we are only pumping on the manifold (with the vacuum vessel valved off). At this point in time, no signal is seen on the leak checker during standard helium leak checking operations.

\begin{figure}[t!]
    \centering
    \includegraphics[width=0.7\textwidth]{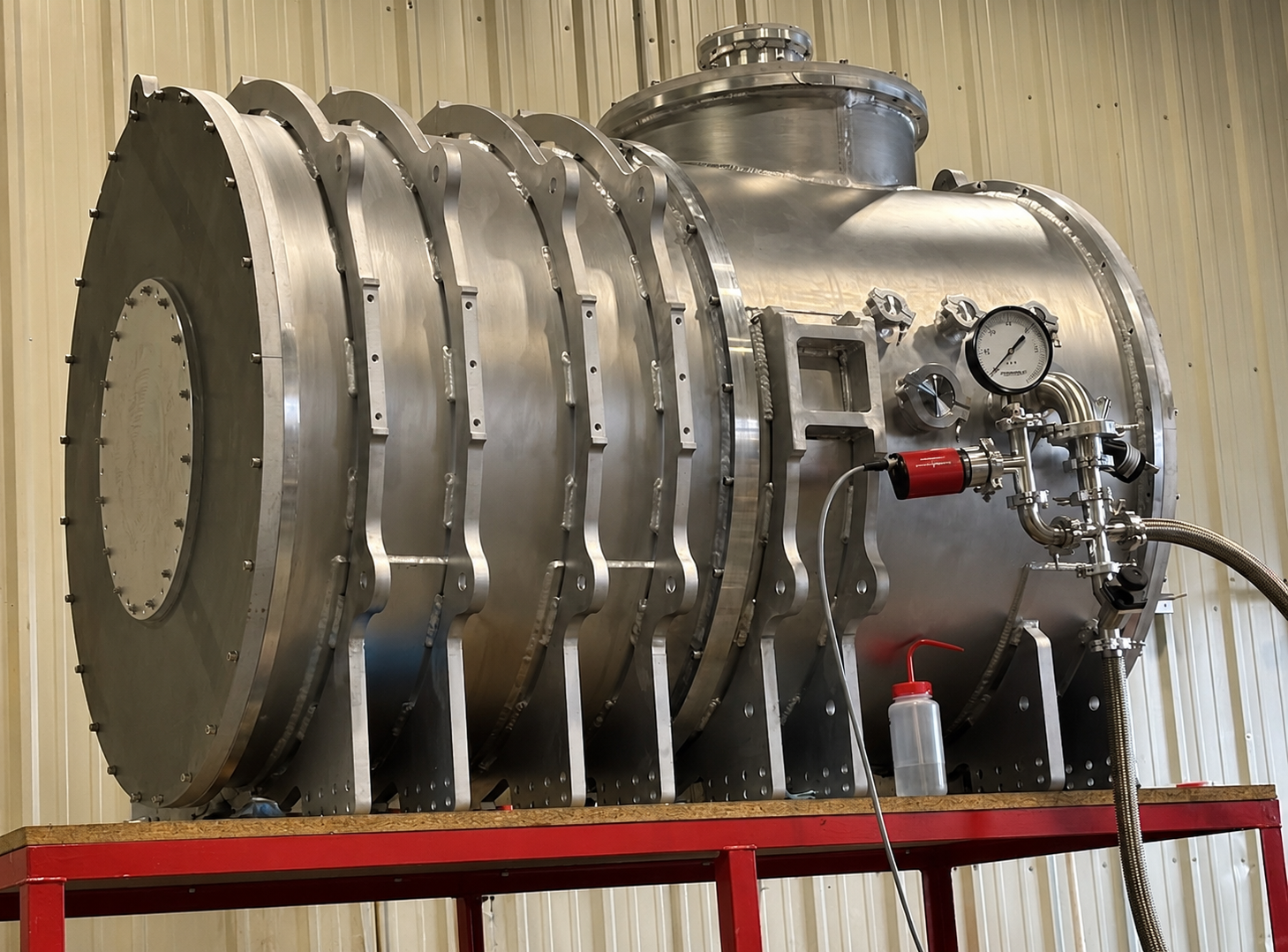}
    \caption{View of the port side of the cryostat vacuum vessel at the conclusion of leak checking operations. The standard pumping and leak checking manifold is shown on the right side of the figure. Two KF25 pumping lines lead to the leak checker and roughing pump.}
    \label{fig:cryostat_finished}
\end{figure}

\section{Summary and Conclusions}

We describe the design, manufacturing, and leak testing of a \SIadj{4}{\kelvin} cryostat that will be used for millimeter-wave holography in the coming years. Room-temperature vacuum qualification reduced the \SIadj{10}{\minute} valve-off pressure-rise rate from approximately \SI{1e-1}{\milli\bar\per\minute} to \SI{3.6e-4}{\milli\bar\per\minute}. After the final repair campaign, standard helium spray testing produced no localized response above the measured background, indicating that the remaining pressure rise was no longer dominated by gross external leaks. Figure \ref{fig:cryostat_finished} shows the assembled vacuum vessel at the conclusion of leak testing at Hedinn. The vacuum vessel has now been shipped to the University of Iceland where the 45- and \SIadj{4}{\kelvin} shells are being wrapped with MLI prior to final assembly. We plan to perform the first cooldown of the cryostat in late August 2026. This is the first cryostat that has been fully designed and built in Iceland.

\acknowledgments % equivalent to \section*{ACKNOWLEDGMENTS}       
 
We kindly thank Ashali Ásrún Gunnarsdóttir, Ashesh Khatua, and Katrín Hekla Magnúsdóttir for their help with the leak checking efforts. We are thankful to Arnar Þórðarson, Jón Trausti Guðmundsson, and Jakob Valgarð Óðinsson from Hedinn for a great collaboration on the design and construction of this cryostat. We also thank Steven Benton, William C. Jones, Johanna Nagy, Michael Niemack, and Simon Tartakovsky for very helpful conversations about the cryostat design. 

Funded in part by the European Union (ERC, CMBeam, 101040169). JEG gratefully acknowledges support from the University of Iceland Research Fund and the Icelandic Research Fund (Grant number: 2410656-051) and the Icelandic Infrastructure fund (Grant number: 263417-901).

% References
\bibliography{report} % bibliography data in report.bib
\bibliographystyle{spiebib} % makes bibtex use spiebib.bst

\end{document}